\documentclass[apjl]{emulateapj}
\usepackage{natbib}
\usepackage{graphics}
\shorttitle{NGC 5128 GCs}
\shortauthors{COLUCCI ET AL.}

\begin{document}

\newcommand{\msol}{M_\odot}
\newcommand{\kms}{km~s$^{-1}$}
\newcommand{\rkms}{km~s$^{-1}$ \enskip}
\newcommand{\rAA}{{\AA \enskip}}
\defcitealias{m31paper}{C09}
\defcitealias{paper3}{C11}
\defcitealias{paper4}{C12}
\defcitealias{mb08}{MB08}

\title{Chemical Abundance Evidence of Enduring High Star Formation Rates  in an Early Type Galaxy:  High [Ca/Fe]  in NGC 5128 Globular Clusters\footnotemark[1]}

\footnotetext[1]{This paper includes data gathered with the 6.5 meter Magellan 
Telescopes located at Las Campanas Observatory, Chile.}

\author{Janet E. Colucci$^{1}$, Mar\'ia Fernanda Dur\'an$^{1}$, Rebecca A. Bernstein$^{1}$,   \& Andrew McWilliam$^{2}$}

\affil{1. Department of Astronomy and Astrophysics, 1156  High Street, UCO/Lick Observatory,\\ University of California, Santa Cruz, CA 95064\\ 2 . The Observatories of the Carnegie Institute of Washington, 813 Santa Barbara Street, Pasadena, CA 91101-1292}
\email{ jcolucci@ucolick.org}

\submitted{Received May 16, 2013, Accepted June 27, 2013}

\begin{abstract}
  We present [Fe/H], ages, and Ca abundances for an initial sample of 10 globular clusters  in NGC 5128 obtained from high resolution, high signal-to-noise ratio  echelle spectra of their integrated light. All abundances and ages are obtained  using our original technique for high resolution integrated light abundance analysis of globular clusters.  The clusters  have a range in [Fe/H] between $-1.6$ to $-0.2$.  
In this  sample, the average [Ca/Fe] for clusters  with [Fe/H]$<-0.4$ is $+$0.37$\pm$0.07, while the average [Ca/Fe] in our MW and M31 GC samples is $+$0.29 $\pm$0.09 and $+$0.24 $\pm$0.10, respectively.   This may imply a more rapid chemical enrichment history for NGC 5128 than for either the Milky Way or M31.This sample provides the first quantitative  picture of the chemical history of NGC 5128 that is directly comparable to what is available for the Milky Way.    Data presented here were obtained with the  MIKE echelle spectrograph on the Magellan Clay Telescope.
\end{abstract}


\section{Introduction}
\label{sec:intro}
\setcounter{footnote}{1}

 For a full understanding of galaxy formation, galaxies of all Hubble types must be studied. 
In this regard, the formation history of  NGC 5128 (Centaurus A)  is especially important because it is the closest  early type galaxy to the Milky Way (MW), and therefore can be observed in the greatest detail. Although it is the closest, the distance to NGC 5128 (3.8 Mpc, \cite{harris10}), makes it impossible to obtain spectra of its individual stars, let alone that required for detailed chemical abundance analysis.

With the development of our original technique for abundance analysis of  integrated light (IL) spectra of globular clusters (GCs), we can now make significant advances in chemical evolution studies of distant massive galaxies.  Because they are luminous and therefore observationally accessible to large distances, unresolved GCs can provide chemical enrichment and formation histories of other galaxies, just as they were originally used to learn about the formation of the Milky Way \citep[e.g.][]{1962ApJ...136..748E,1978ApJ...225..357S}.  Our technique has been developed and demonstrated on resolved GCs in the Milky Way and Large Magellanic Cloud (LMC) in  \cite{bernstein05}, \cite{mb08} (hereafter ``MB08''),\cite{milkyway}, \cite{paper3} (hereafter ``C11''), \cite{paper4} (hereafter ``C12"), and \cite{stars}.   These publications  demonstrate that the IL analysis provides  Fe abundances and [X/Fe] ratios accurate to $\sim$0.1 dex, as well as  distinguishes ages  for GCs with [Fe/H] of $-2$ to $+0$ and ages in the range of  0.05 to 12 Gyr. We also note the application of an independent, but  similar technique by \cite{larsen} to GCs in the Fornax dwarf spheroidal galaxy.

Using this method, we have now  begun a  study of the chemical composition of the GC system of NGC 5128.  Here we present Fe abundances, ages, and Ca abundances of  10 GCs in our  initial sample. Our complete sample will include approximately 20 GCs, to provide better statistics for galaxy to galaxy comparisons. In  future  papers  we will present detailed abundances of an additional $\sim$20 elements in our complete sample of 20 GCs.

In \textsection \ref{sec:obs}, we describe the target selection, observations and data reduction. Our abundance analysis is briefly described in \textsection \ref{sec: abund} and our results in \textsection \ref{sec: results}.

\section{Targets, Observations and Reductions}
\label{sec:obs}

For the targets observed prior to 2010, we chose GCs from the catalog of  \cite{pengcat} that were more luminous than V$\sim$18.5 mag, that are in relatively uncrowded regions, that are not projected onto the high surface brightness part of the NGC 5128 bulge, and that have a range in projected galactocentric distance from NGC 5128.  For the targets observed after 2010, we  also aimed to observe GCs with as wide a range in estimated metallicity and age as possible, using the additional metallicity and age  estimates that became available in \cite{beasley08} and \cite{woodleymet}.
While our sample is obviously not complete, our attempt to select   GCs with a wide range in previously estimated [Fe/H], age,  and   projected galactocentric distance from NGC 5128 (R$_{\rm gc}$) increases our chances of surveying the range of properties present in NGC 5128 GCs.  The 
magnitudes, spatial information and previously measured properties are listed for all of the GCs  
in Table~\ref{tab:obs}.   We list both the names from \cite{pengcat} and those from the newer, homogenized catalog of \cite{woodley07}.  We use the GC names of \cite{woodley07} for the remainder of this work.

We obtained our spectra  using the  MIKE spectrograph
 \citep{mike} on the 6.5m Magellan Clay telescope. The data were taken over six observing runs from 2004-2012.  All data were taken with a slit size of 1x5 arcsec, providing an instrumental resolution of 28,000 for the blue arm and 22,000 for the red arm. 
 The wavelength coverage of the blue and red arms of MIKE  are  approximately 3350-5050 \AA~ and 4800-9000 \AA , respectively.
Exposure times for this sample are between 5$-$17 hours for each GC
and are listed in Table~\ref{tab:obs} along with the date each GC was observed. For all GCs, with the exception of GC0106, we  only use exposures where the seeing was $\leq$1.0''; for GC0106  25\% of the exposures were taken with seeing between 1.0''-1.3''. Signal-to-noise (SNR)
estimates at 6040 \rAA are  given in Table~\ref{tab:obs}. 
 Data were reduced with the
MIKE Redux pipeline.\footnote{ http://www.ucolick.org/~xavier/IDL/index.html} 
A heliocentric velocity correction was applied and final radial velocities (${\rm v}_{r}$) were measured with the IRAF task {\it rvidlines}. One dimensional, line of sight velocity dispersions (${\rm v}_{\sigma}$) were measured using the IRAF task {\it fxcor}  and red giant template stars that were observed with identical setups as the GCs. These measurements are discussed briefly in \textsection \ref{sec: results}.

\section{Abundance Analysis}
\label{sec: abund}

Fe I and Ca I  lines and line parameters  were taken from our previous spectroscopic analyses 
\citepalias[][and references therein]{mb08, m31paper, paper4}.  The abundances from the Fe I and Ca I  lines 
that are included in our final analysis are listed in Table
~\ref{tab:linetable}. 
As in all of our previous spectroscopic analyses, we use the  ODFNEW and AODFNEW  model stellar atmospheres from Kurucz\footnote{http://kurucz.harvard.edu/grids.html} 
\citep{2004astro.ph..5087C}. All abundances are calculated under the assumption of local
thermodynamic equilibrium (LTE), and abundance ratios relative to the sun were calculated with the solar abundance values of \cite{asplund09}.

Our IL abundance analysis methods are  described in detail in \citetalias{mb08}, \citetalias{m31paper},  \cite{milkyway}, \citetalias{paper3} and \citetalias{paper4}. Briefly, we create synthetic GC color magnitude diagrams (CMDs)  using the Teramo isochrones \citep{2004ApJ...612..168P,2006ApJ...642..797P} without convective overshooting,  with extended asymptotic giant
branch, $\alpha-$enhanced low$-$temperature opacities
calculated according to \cite{ 2005ApJ...623..585F}, and mass$-$loss parameter of $\eta$=0.2.  More discussion about the testing and choice of appropriate isochrones can be found in \citetalias{mb08}, \citetalias{paper3} and \cite{stars}.

To measure abundances, we perform a flux-weighted spectral synthesis for a  $\sim$20 \AA~ region around each line using
our routine ILABUNDS \citepalias[see][]{mb08,paper4}, which utilizes 
spectral synthesis routines from the 2010 version of MOOG \citep{1973ApJ...184..839S}.  
We then find the abundance that produces the best matching synthetic spectra using a $\chi^{2}$-minimization scheme, as described in \citetalias{paper4}.
Complete synthesis of a large region allows for more accurate continuum placement and a more accurate appraisal of line blending then direct measurement of equivalent widths (EWs) by gaussian fitting. To supplement our select line list of Fe I and Ca I, we use a combination of  VALD and Kurucz  line lists;  for $\lambda>$6300 \rAA these lists  have been calibrated to both the Sun and Arcturus by \cite{evan1}.   The final synthesized spectra were convolved with the  ${\rm v}_{\sigma}$ measured for each GC. The final age and [Fe/H]  solutions for  each GC are  identified as the range in synthetic CMD ages and [Fe/H] that produce the most self-consistent results using the 20$-$40 individual Fe I lines measured in each cluster.  
The best solutions have the smallest statistical error ($\sigma_{{\rm N}}$; the standard deviation of the mean abundance), and minimal dependence of Fe I abundance with line excitation potential (EP), wavelength ($\lambda$), and EW.  Results for the NGC 5128 sample are given in Table \ref{tab:abund}.

For each GC there is a range in CMD ages that produce similarly self-consistent solutions.  For older GCs this range is typically 10$-$15 Gyr, and leads to  a systematic uncertainty in [Fe/H] of $\lesssim$ 0.05 dex, which we denote $\sigma_{\rm {Age}}$.  For GCs with a larger acceptable age range, the $\sigma_{\rm {Age}}$  is correspondingly larger ($\sim$ 0.1 dex).   For the total uncertainty in [Fe/H] for each cluster ($\sigma_{\rm{T}}$), we add the statistical error in the mean abundance ($\sigma_{ {\rm N}} / \sqrt{{\rm N_{Fe}} -1} $) and the systematic age uncertainty $\sigma_{\rm {Age}}$ in quadrature, all of which are listed in Table \ref{tab:abund}.

\begin{figure}
\centering
\includegraphics[trim= 0mm 0mm 30mm 0mm, scale=0.50]{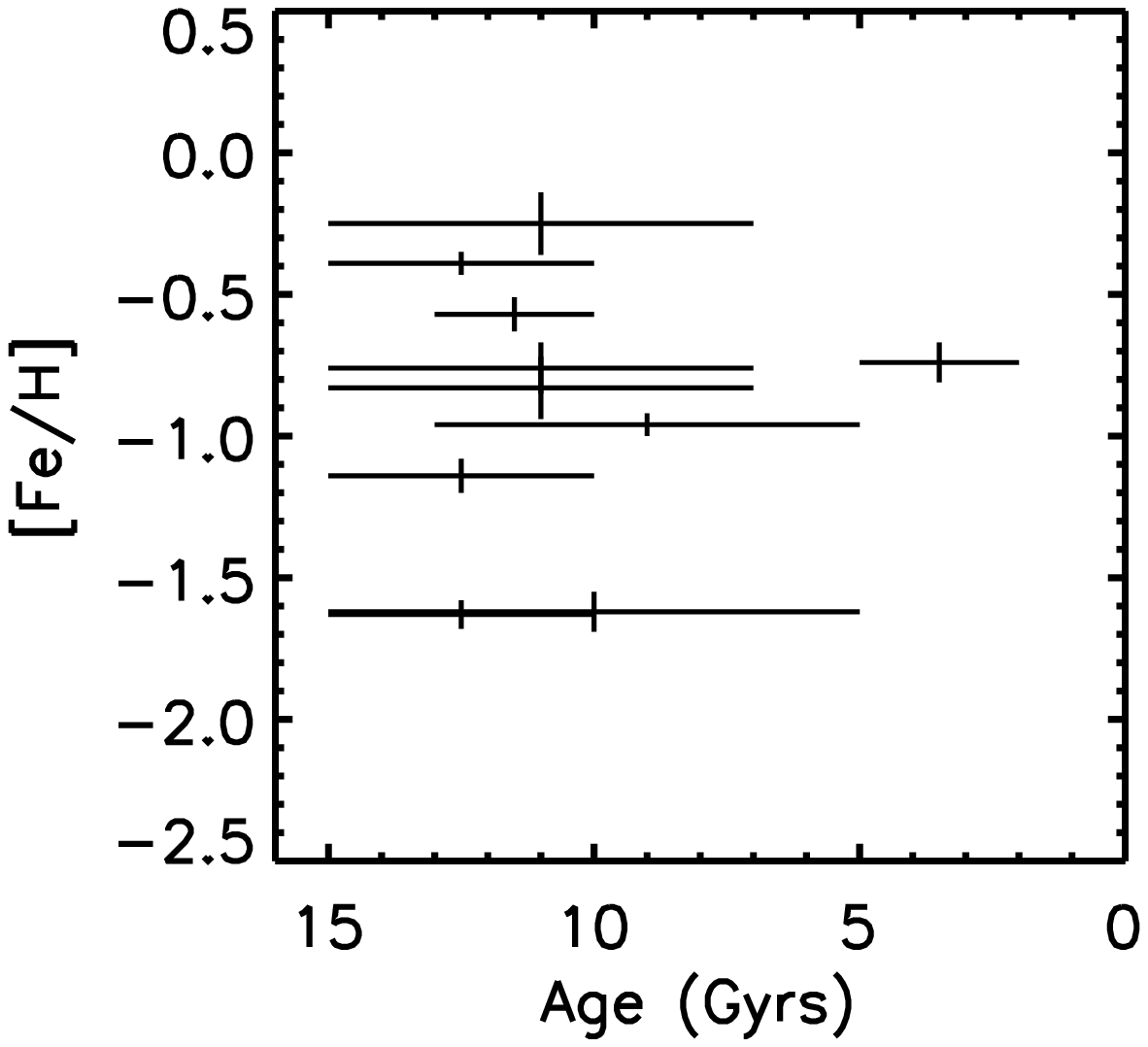}
\includegraphics[trim= 10mm 0mm 60mm 0mm, scale=0.50]{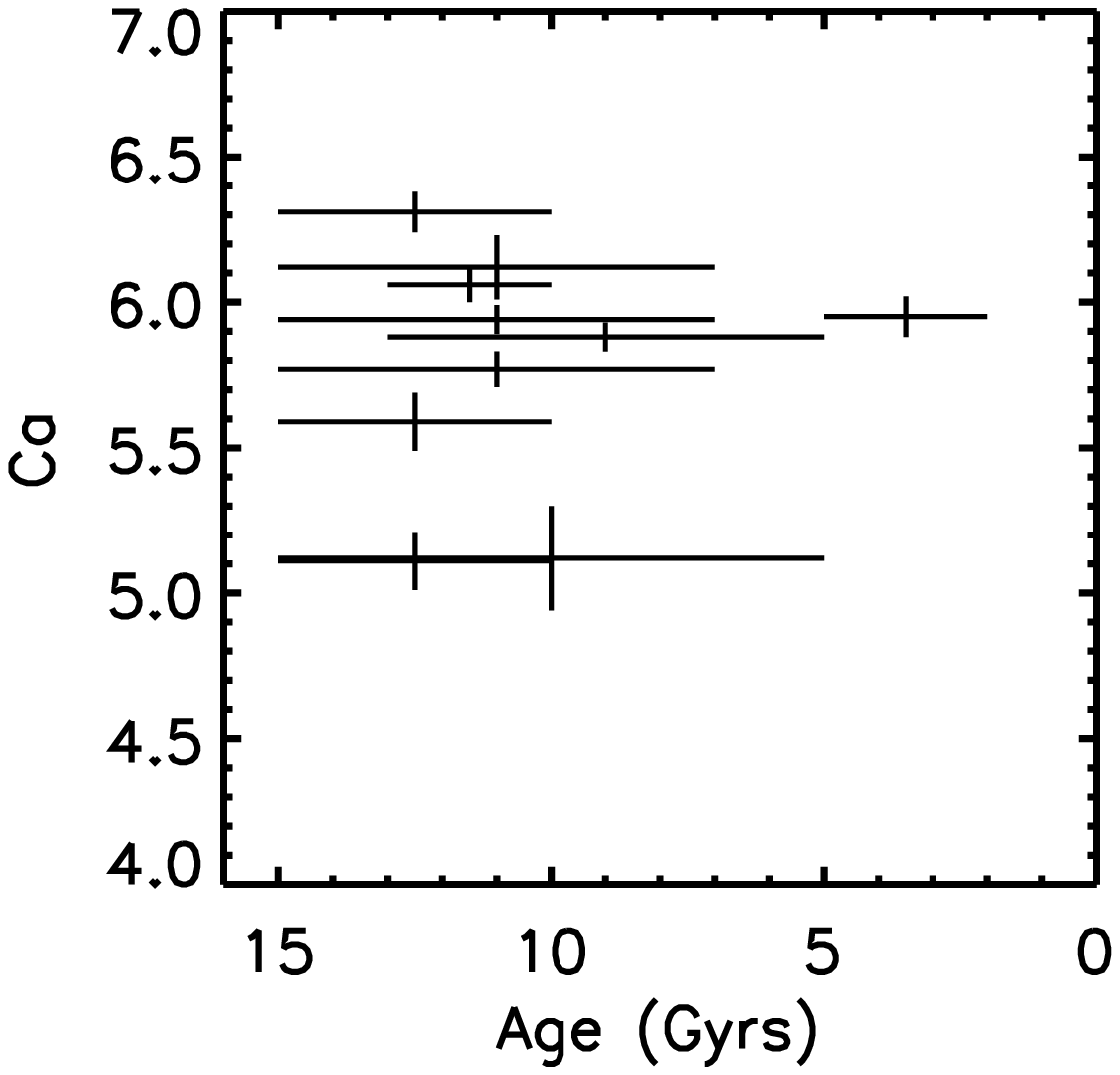}
\includegraphics[trim= 0mm 0mm 30mm 0mm,scale=0.50]{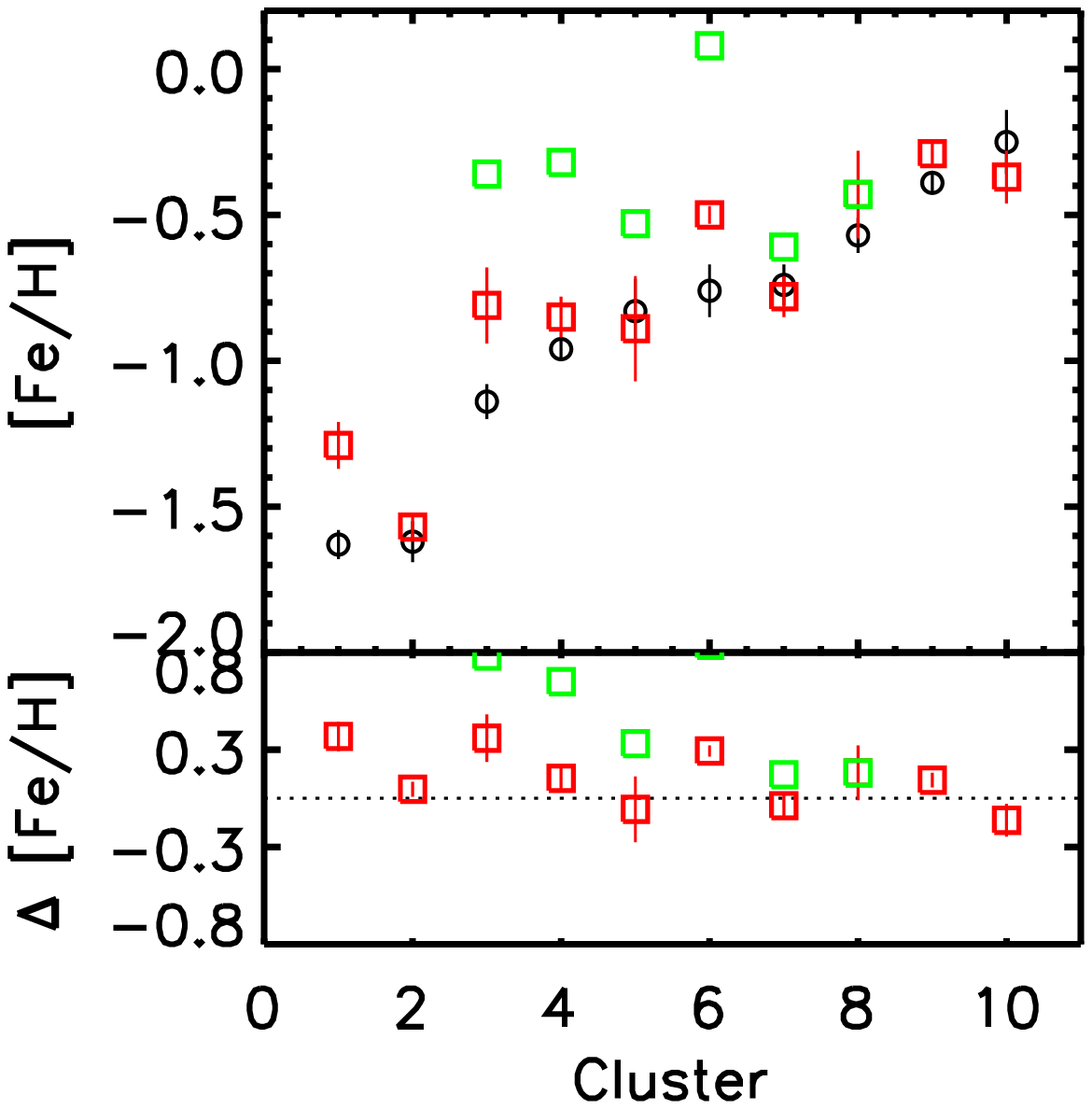}
\caption{Fe, Ca, and age results.  The results for NGC 5128 GCs are shown in the top  panels.  In the bottom  panel,  
we show comparisons of  ``metallicity''  from different techniques;  the red squares show the [M/H] of \cite{beasley08}, green squares show the [Z/H] of \cite{woodleymet}, and black circles show the [Fe/H] of our analysis.
GCs have been sorted  according to increasing [Fe/H] from our analysis. The corresponding Y-axis cluster designations 1-10 are GC0227, GC0106, GC0277, GC0365, GC0282, GC086, GC0265, GC0124, GC0041, and GC0040, respectively. The lower section of this panel shows the difference in metallicity between our results and those of \cite{beasley08} and \cite{woodleymet}.}
\label{fig:compare} 
\end{figure}

\begin{figure*}
\centering
\includegraphics[scale=0.7]{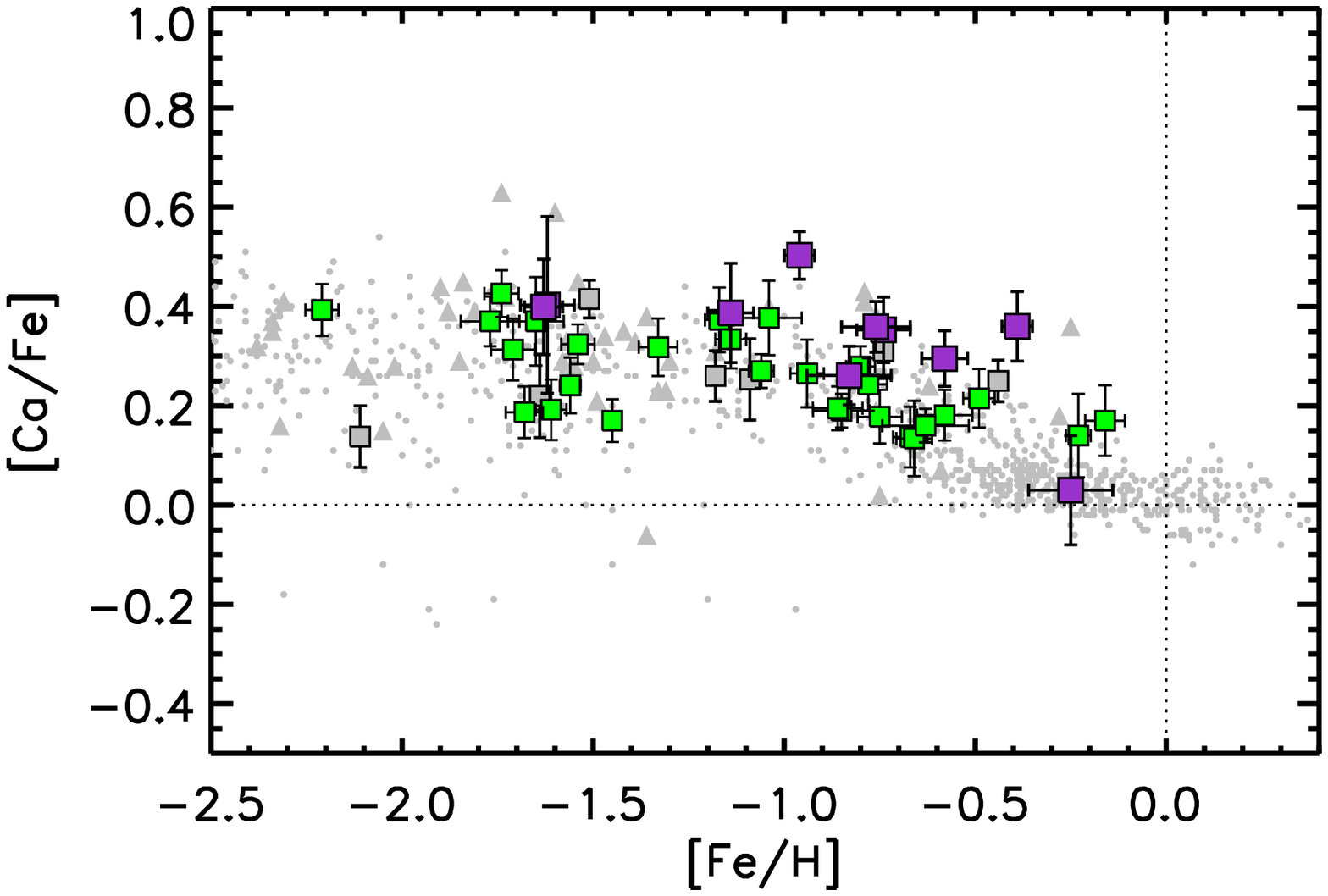}
\includegraphics[scale=0.7]{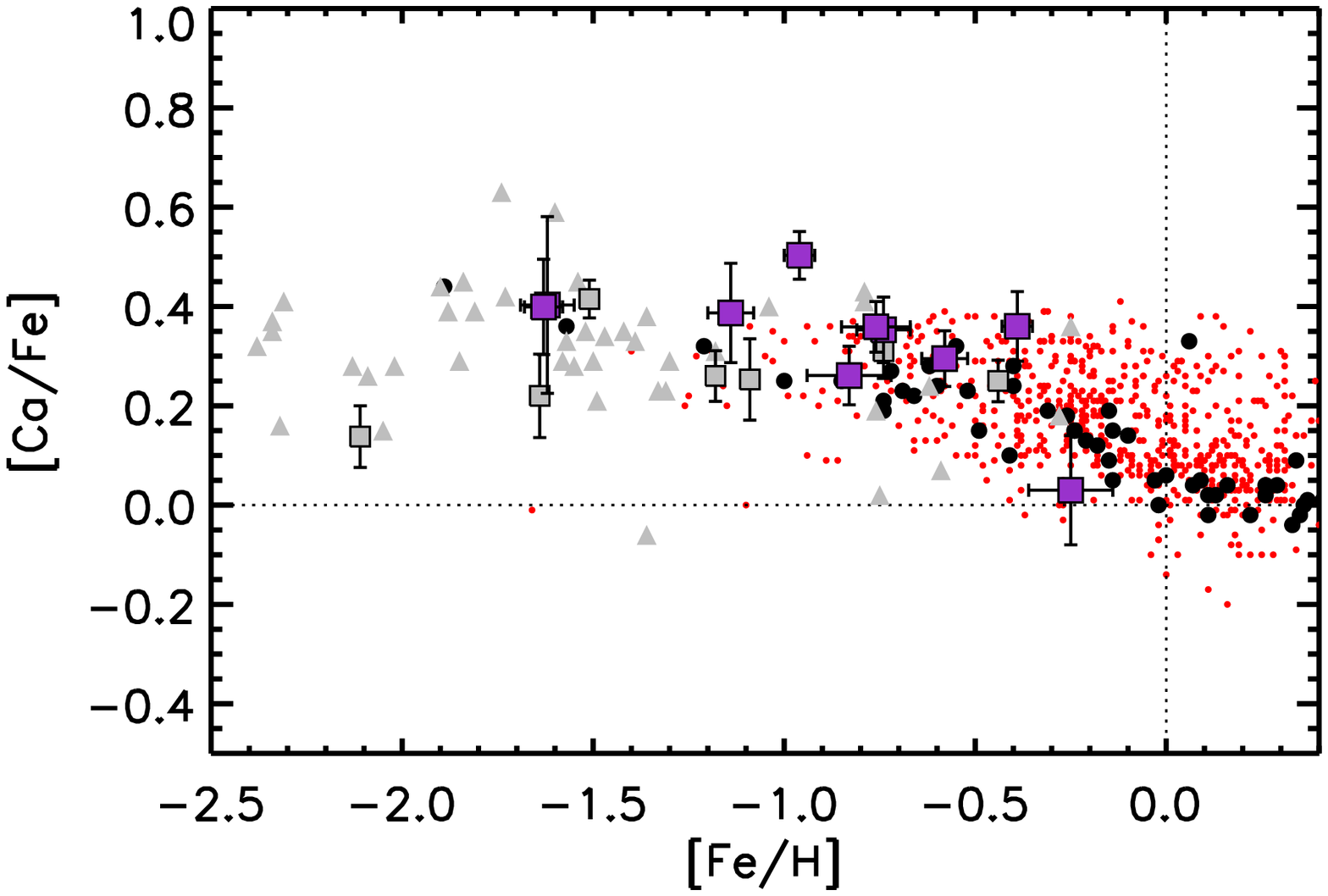}
\caption{
The top panel shows [Ca/Fe] results for NGC 5128 GCs  (purple squares) compared with MW halo and disk field stars (small gray circles, \cite{venn})   stellar GC abundances (small gray triangles,  \cite{pritzl}),  MW IL abundances (gray squares, \cite{milkyway}) and M31 GCs (green squares,  \cite{rome}).  In the bottom panel the GCs in NGC 5128 and the MW are shown with MW bulge stars only.  The bulge samples of \cite{gonzalez11}, and  \cite{bensby} correspond to red and black circles, respectively.
}
\label{fig:feh} 
\end{figure*}

\section{Results and Discussion}
\label{sec: results}
Final [Fe/H] abundances, ages and  [Ca/Fe] abundances are reported in Table 3, along with our measured radial velocities (v$_{r}$) and velocity dispersions  (${\rm v}_{\sigma}$). All of the previous measurements of   ${\rm v}_{\sigma}$ and  v$_{r}$  listed in Table 1, which were obtained with lower resolution or lower SNR spectra, agree with our results to within  3 $\sigma$,  with  most agreeing to within 1 $\sigma$. We also note that differences in the observed velocity dispersions with respect to those of \cite{taylor10} could be attributed to aperture effects, the seeing during our observations, and their impact on the fraction of the light of the GC in the spectrograph slit;   however quantification of these effects is not necessary for the abundance analysis, and therefore is  beyond the scope of this work.

In Figure  \ref{fig:compare}, we show our [Fe/H] and Ca  results as a function of age.  The  [Fe/H] for our current sample ranges from $-1.6$ to $-0.25$, which  is well within the range exhibited by MW GCs \citep{harriscat}.    In the bottom panel of Figure \ref{fig:compare}, we compare the metallicity estimates of \cite{beasley08} and \cite{woodleymet} to our [Fe/H] measurements, although we note that metallicities from low resolution line index techniques do not strictly measure [Fe/H].  Nevertheless, the comparison of spectroscopic metallicity scales is interesting because of the large samples that can be efficiently obtained at low spectral resolutions.
 Figure \ref{fig:compare} shows that the [Z/H] of \cite{woodleymet}, which only cover the range $-1 <$ [Fe/H] $< 0$ in this sample, are biased high by as much as 1.0 dex. The [M/H] of \cite{beasley08}, which can be compared over the full [Fe/H] range of our sample, are more consistent with our results and have a scatter of $\sim$0.3 dex.

 Nearly all  of the GCs have ages that are consistent  with an ancient population ($>$10 Gyr), to within the uncertainties. In one case, GC0265,  we measure an age of 2-5 Gyr.  This GC has one previous age estimate by \cite{woodleymet} of 8.5 $\pm$ 0.1 Gyr.  While the value of \cite{woodleymet} is not  consistent on an absolute age scale,  it is consistent in the general sense that this GC has a younger age than a typical $>$10 Gyr ancient population. It is well known that blue horizontal branch (HB) stars can mimic the presence of truly young stars in IL. Therefore, in our analysis, we have additionally verified that it is not possible to obtain a better, more self-consistent solution for GC0265 with a synthetic CMD that is $>$10 Gyr, and has a very blue HB.   This is important to test explicitly because  the Teramo isochrones are parametrized such that they do not produce CMDs with blue HBs for [Fe/H]$\sim$-0.7.  Our technique for evaluating the influence of blue HB stars is described in Colucci et al. (2013, in prep) along with  quantitative constraints of HB morphology in a large sample of unresolved GCs.

 For the other 5 GCs that we have in common with \cite{woodleymet},   those authors measure intermediate ages for 3 of the GCs, while we find old ages, as stated above. In at least two of these cases our analysis is consistent with these GCs hosting a significant number of blue HB stars, which would cause the young age measurements in the \cite{woodleymet} analysis.  An interesting test would be to compare the GALEX FUV measurements for these GCs to confirm our identification of blue HB content as opposed to truly young populations. \cite{reyUV} used GALEX observations of NGC 5128 GCs to infer that there is indeed a significant intermediate age GC population in NGC 5128, but did not include the full catalog of individual GCs, so we unfortunately cannot cross check results for our sample of GCs at this time.

As seen in Figure \ref{fig:feh}, nearly all of the GCs in our initial sample are enhanced in [Ca/Fe] relative to the Sun. [Ca/Fe] is our best proxy for [$\alpha$/Fe], because [Mg/Fe] and [O/Fe] suffer from intracluster abundance variations \citep{m31paper} and because Si and Ti are more difficult to measure. We will present Si and Ti abundances  for a subset of the GCs in an upcoming paper.
 We emphasize that this is the first time that Figure \ref{fig:feh}, with precise abundances that are directly comparable to MW abundances,  can be made over a wide range of [Fe/H] for the oldest stellar populations in {\it any} early type galaxy.  
  For comparison,  we also show the GC IL abundance results for M31 GCs from \cite{rome}, which closely track the abundances in MW disk stars and  thus show similar chemical enrichment histories in large spirals in the Local Group.  
Visual inspection of the top  panel  of Figure \ref{fig:feh} shows that for the  NGC 5128 GCs, the mean [Ca/Fe] in GCs with  [Fe/H] $< -0.4$ appears higher than for the GCs in M31 at the same [Fe/H], and higher than MW GCs and disk stars.   Because our initial sample is small, we do not perform a quantitative statistical test to compare the different galaxies in Figure \ref{fig:feh}; this test will be performed with our larger sample of 20 total GCs in an upcoming paper.   For the purposes of this paper, we calculate a simple mean of the NGC 5128 sample (for [Fe/H]$<-0.4$), which gives [Ca/Fe]=$+0.37 \pm 0.07$, where the uncertainty is the standard deviation of the mean.  This is indeed higher than  the average values for the MW and M31 GC samples  in the same metallicity range, which are $+0.29 \pm$0.09 and $+0.24 \pm$0.10, respectively.
This may imply that NGC 5128 experienced a more prolonged  high star formation rate era, with faster chemical enrichment, than either M31 or the MW.  If the knee in [Ca/Fe] is at higher [Fe/H] than the MW disk, it more closely resembles recent results for MW bulge stars by \cite{gonzalez11} and \cite{bensby}, which would imply similar formation histories for spheroids.  We compare the GC abundances to the bulge samples in the bottom panel of Figure \ref{fig:feh}, where it can be seen that there is more overlap with the GCs in [Ca/Fe] at high [Fe/H] for the bulge stars than for the disk stars in the top panel.

Our initial sample of GC abundance measurements in NGC 5128 support a scenario where most of the halo formed at very early times, with rapid chemical enrichment reaching to high overall [Fe/H], such as that discussed in \cite{rejkuba11}. From  deep HST/ACS photometry of halo stars, \cite{rejkuba11} find that the observed CMDs are most consistent with a population where 70-80\%  of the stars are ancient,  reach super-solar [Fe/H], and are  $\alpha$-enhanced, which is similar to our results for GCs. \cite{rejkuba11} also find that their data suggest that the  younger   population in the  halo formed in a second burst of star formation 2-4 Gyr ago. This is interesting in the context  of our GC measurements, because we find one GC  with  a younger age that is consistent with the age of the proposed burst. In addition,  this GC  has an enhanced [Ca/Fe] ratio, which is more consistent with a burst of star formation than with extended star formation. It's also worth noting that  [O/H] measurements in the halo planetary nebulae of NGC 5128 also support a recent burst of star formation that formed stars reaching to high $\alpha$-enhancement \citep{walsh12}.

 Finally we note that higher [$\alpha$/Fe] in NGC 5128 than in the MW is not the same conclusion reached by groups using low resolution spectra; \cite{pengmet} estimated that NGC 5128 GCs have similar [$\alpha$/Fe]  to the  MW, and  both \cite{beasley08} and \cite{woodleymet} concluded that the mean [$\alpha$/Fe] in NGC 5128 GCs is lower than the MW.  This is not surprising given that [$\alpha$/Fe] estimates from line indexes have large uncertainties and are plagued by systematics.

\section{Summary}
\label{sec:sum}

We have measured ages and  abundances for Fe and Ca for an initial sample of 10 GCs in NGC 5128 using an original technique for detailed abundance analysis of high resolution integrated light spectra. We find that the majority of GCs in the sample  have old ages ($>$10 Gyr), with only one GC in the initial sample showing  an intermediate age of 2-5 Gyr.  The GCs have a range in [Fe/H] between $-1.6$ to $-0.2$.   The average [Ca/Fe] for GCs with [Fe/H]$<-0.4$ is   higher than for M31 GCs or MW GCs and disk stars, which may imply a more prolonged high epoch of high star formation rate in  NGC 5128 than for either the MW or M31.  Analysis of an additional $\sim$20 elements in our full  sample of  20 GCs will be presented in an upcoming paper.

\acknowledgements
This research was supported by NSF grant AST-0507350. MFD acknowledges support from a Fulbright-Conicyt scholarship. The authors thank the anonymous referee for thoughtful comments that improved this paper.

\bibliographystyle{apj}

\clearpage

{\scriptsize
\begin{table}[tp]\centering
\begin{tabular}{llcccrllllc|lrc}

& & & & & &   \\
\hline
\hline

 Name  & Name  & RA &  Dec&   V  &  ${\rm R}_{{\rm gc}}$ &[M/H]$^{a}$ & [Z/H]$^{b}$ & [$\alpha$/Fe]$^{b}$ &v$_{r}$ & {\rm v}$_{\sigma}$$^{c}$  & Date$^{d}$  & T &  SNR \\ 
P04 & W07 &(J2000) &  (J2000) & & (kpc)&   &&& (km/s)&(km/s)&  &(h) &  (6040 \AA)  \\
\hline

  HGHH-41&GC0040  &    13:24:39.0 &    -43:20:06.4 & 18.59  &23.1&$-$0.37 & \nodata & \nodata & 363$\pm$1 & 13.7$\pm$1.3              & 05May       &    16.6      &  81        \\

 HGHH-29& GC0041  &    13:24:40.4 &    -43:18:05.3 & 18.15  & 21.0  &$-$0.29& \nodata & \nodata  &726$\pm$1  & 17.6$\pm$1.8  &    04Jan       &    9.8     &  64       \\

  HGHH-31 & GC0086  &    13:24:57.4 &    -43:01:08.1 & 18.38  & 6.1 &$-$0.50 & $+$0.08 & $+$0.07&   690$\pm$18 &  \nodata     & 11Apr       &    12.0      &   74       \\

 HGHH-4& GC0106  &    13:25:01.8 &    -43:09:25.4 & 18.04& 10.5  &$-$1.57& \nodata & \nodata  & 689$\pm$16 & \nodata &04Jun       &    15.0      &  73        \\
 
 HGHH-G342 & GC0124  &    13:25:05.8 &    -42:59:00.6 & 18.18  & 5.0  &$-$0.43 &  $-$0.43&$+$0.25& 553$\pm$21  & \nodata & 12Feb       &    11.5      &    66      \\

 HGHH-44 & GC0227  &    13:25:31.7 &    -43:19:22.6 & 18.69  & 20.2  &$-$1.29&\nodata &\nodata & 505$\pm$1 & 13.6$\pm$1.2    & 11Apr       &    14.0      &    68      \\

 HGHH-17 & GC0265 &    13:25:39.7 &    -42:55:59.2 & 17.63 &  6.2 &$-$0.78&  $-$0.61 & $+$0.08    &  782$\pm$2   &20.8$\pm$2.9   & 04Jan      &    10.9     &  80       \\
&&&&&&&&&&& 05May\\

 HGHH-G370 & GC0277  &    13:25:42.3 &    -42:59:17.0 & 18.39   & 3.6 &$-$0.81 & $-$0.36 & $-$0.07 &  507$\pm$34  & \nodata & 12May       &    6.5      &     38    \\

 HGHH-19 & GC0282 &    13:25:43.4 &    -43:07:22.8 & 18.12  &7.6 &$-$0.89 & $-$0.53 & $+$0.25& 632$\pm$10  & \nodata & 12May       &    6.2      &   69       \\

  HGHH-7& GC0365  &    13:26:05.4 &    -42:56:32.4 & 17.17  & 9.2 & $-$0.85 & $-$0.32 &$+$0.01& 595$\pm$1 &23.7$\pm$2.4   & 11Apr       &    5.0      &    71    \\

 \hline

\end{tabular}
\centering
\caption{ Observations and GC properties from the Literature \label{tab:obs} }
\tablecomments{Column 1 names, coordinates, and V magnitudes  are taken from \cite{pengcat}. Column 2 names, projected galactocentric radii from NGC 5128,  and radial velocities (v$_{r}$) are taken  from  \cite{woodley07}. Note that the  v$_{r}$ from \cite{woodley07} are averages from a compilation of velocity measurements from spectra of different resolution and SNR.   ${\rm R}_{{\rm gc}}$ were calculated assuming a distance to NGC 5128 of 3.8 Mpc \citep{harris10}.  (a)  The [M/H]  metallicity value of \cite{beasley08}. (b) The [Z/H] metallicity value and [$\alpha$/Fe] estimate by \cite{woodleymet}.  (c) The line of sight velocity dispersion ({\rm v}$_{\sigma}$) measured by  \cite{taylor10}. (d) The Month and year of the multi-night observing runs is listed in the format YYMonth. }
\end{table}

}

\begin{table}[tp]\centering
\begin{tabular}{lccr|cccccccccc}
& & & \\
\hline
\hline
 &     &   &    & \multicolumn{10}{c}{Abundance}\\
Species& $\lambda$  & EP &  log gf &  GC0040 & GC0041 & GC0086 & GC0106 & GC0124 & GC0227 &GC0265 & GC0277 & GC0282 & GC0365 \\
\hline
\hline
    Fe  I &5497.526 &  1.010 & -2.849 & 7.12 & 7.31 & \nodata & \nodata & \nodata & \nodata & 6.50 & 6.41 & \nodata & \nodata \\
 Fe  I &5501.477 &  0.958 & -3.047 & 7.22 & \nodata & 6.32 & \nodata & 6.93 & 5.51 & \nodata & 6.31 & 6.94 & 6.39 \\
 Fe  I &5506.791 &  0.990 & -2.797 & \nodata & \nodata & 6.52 & \nodata & 6.73 & 5.81 & 6.70 & 6.31 & 6.64 & 6.69 \\
 Fe  I &5569.631 &  3.417 & -0.486 & 7.22 & \nodata & 6.72 & \nodata & \nodata & \nodata & 6.70 & \nodata & 6.74 & \nodata \\
 Fe  I &5576.099 &  3.430 & -0.900 & \nodata & 6.91 & 6.82 & \nodata & 7.23 & 6.01 & \nodata & \nodata & 6.74 & 6.79 \\
 \nodata \\
 \\

 Ca  I &6166.440 &  2.520 & -1.142 & \nodata & 6.54 & 6.13 & \nodata & 5.94 & \nodata & \nodata & \nodata & 5.65 & 5.90 \\
 Ca  I &6169.044 &  2.520 & -0.797 & \nodata & 6.54 & 5.83 & \nodata & 6.04 & 5.12 & \nodata & 5.62 & 5.75 & 5.70 \\
 Ca  I &6439.083 &  2.526 &  0.390 & \nodata & 6.24 & 5.93 & 4.87 & 5.84 & 4.72 & 5.92 & 5.82 & 5.75 & 5.70 \\

\nodata \\
  \hline
\end{tabular}
\caption{~Line Parameters and GC IL abundances\label{tab:linetable}. }
\tablecomments{~ Table 2 is presented in its entirety in machine-readable format in the electronic edition. }
\end{table}

\begin{table}[tp]\centering
\begin{tabular}{lrrrrrrrcrrrcc}

& & & \\
\hline

 Name   &  Age & [Fe/H] &  {\rm N$_{\rm Fe}$} &  $\sigma_{ {\rm N}}^{a}$& $\sigma$ &$\sigma_{ {\rm Age}}$ &  $\sigma_{ {\rm T}}^{b}$ &  $\alpha^{c}$ & [Ca/Fe] & {\rm N$_{\rm Ca}$} & $\sigma_{{\rm Ca}}^{d}$   & ${\rm v}_{\sigma}^{e}$ & v$_{r}^{f}$ \\ 
  &  (Gyrs) &  &&&&&&&&& &(\kms) & (\kms) \\

\hline
\hline
 
 GC0040  & 7-15  &  $-$0.25 & 17 & 0.20& 0.05   &0.10 &0.11 & S &  $+$0.03 & 4 & 0.11     &  12.5$\pm$0.24 &   363.3$\pm$0.31\\
 
 GC0041  & 10-15 &  $-$0.39 &  22 & 0.16& 0.03   & 0.03 & 0.04 & A &$+$0.36 & 9 & 0.07  &  16.8$\pm$0.44 &   725.1$\pm$0.24\\
 
 GC0086 & 7-15  & $-$0.76 &  33 & 0.18 &    0.03  & 0.09 & 0.09 & A &  $+$0.36 & 9  & 0.05  & 17.5$\pm$1.09   &    686.8$\pm$1.52 \\
 
 GC0106 & 5-15 & $-$1.62& 21 & 0.18 & 0.04   &0.06 & 0.07 & A &  $+$0.40 & 3 & 0.18   & 14.0$\pm$0.46 &    708.3$\pm$0.35\\
 
 GC0124&  10-13 &  $-$0.57&  31 & 0.22 & 0.04     &0.04& 0.06 & A &$+$0.30 & 8 &  0.06 & 14.7$\pm$0.35 &    523.1$\pm$0.89\\
 
  GC0227& 10-15  &$-$1.63 & 30 & 0.25 &   0.05     &0.05 & 0.05 & A &$+$0.40 & 7 & 0.10  &10.2$\pm$0.79     & 501.6$\pm$0.66 \\

 GC0265 & 2-5 & $-$0.74 & 26 & 0.14 &  0.03      &0.06 & 0.07 & A & $+$0.35 & 6 & 0.07 & 19.1$\pm$0.28&  783.7$\pm$0.38 \\

 GC0277 &10-15 &  $-$1.14& 16 &   0.04     &0.14 & 0.04 & 0.06& A & $+$0.39     & 6   & 0.10   &   15.3$\pm$0.77 & 510.0$\pm$0.97 \\
 
 GC0282 & 7-15 &  $-$0.83 & 37 & 0.23 &  0.04    & 0.10 & 0.11 & S & $+$0.26 & 11 & 0.06  &14.5$\pm$0.34  & 618.3$\pm$0.63 \\
 
 GC0365 &5-13 & $-$0.96 & 31 & 0.17 &    0.03     &0.02 & 0.04 & A & $+$0.50 & 9 & 0.05&  20.9$\pm$1.28  & 597.4$\pm$0.64\\

\hline

\end{tabular}
\caption{NGC 5128 GC Ages and Abundances\label{tab:abund}}
\tablecomments{  ~ 
(a) $\sigma_{ {\rm N}}$ is the standard deviation of the mean abundance.
 (b) The total uncertainty on the [Fe/H] is defined as
  $\sigma_{ {\rm T}}$=$\sqrt{ \sigma^{2} + \sigma_{ {\rm Age}}^{2}  }$, where $\sigma$= $\sigma_{ {\rm N}} / \sqrt{{\rm N_{Fe}} -1} $ is the
  error in the mean abundance and N$_{\rm Fe}$
  is the number of Fe I lines analyzed in each GC. 
(c) A or S designates whether Kurucz $\alpha$-enhanced
  or scaled-solar atmospheres were used, respectively. This decision is based explicitly on the [Ca/Fe]  abundances measured for each GC; a proxy for [$\alpha$/Fe].
  (d) $\sigma_{{\rm Ca}}$ is  the error in the mean measured Ca I abundance. 
  (e) ${\rm v}_{\sigma}$ is the line of sight velocity dispersion of the GC measured by {\it fxcor}, and used to convolve with synthesized spectra.  
  (f) Heliocentric corrected radial velocities. }
\end{table}

\clearpage

\end{document}